\documentclass[manuscript]{aastex62}
\usepackage{amsmath}
\graphicspath{{./}{figures/}}

\shorttitle{New Constraints on CNEOS 2014-01-08}
\shortauthors{Siraj \& Loeb}
\begin{document}

\title{New Constraints on the Composition and Initial Speed of CNEOS 2014-01-08}

\email{amir.siraj@cfa.harvard.edu, aloeb@cfa.harvard.edu}

\author{Amir Siraj}
\affil{Department of Astronomy, Harvard University, 60 Garden Street, Cambridge, MA 02138, USA}

\author{Abraham Loeb}
\affiliation{Department of Astronomy, Harvard University, 60 Garden Street, Cambridge, MA 02138, USA}

%\keywords{asteroids: individual (A/2017 U1)}

%% depreciated in this version as it is no longer . AASTeX 

\begin{abstract}
We study the newly released light curve from the fireball of the first interstellar meteor CNEOS 2014-01-08. The measured velocity and three observed flares down to an altitude of $18.7 \mathrm{\; km}$ imply ambient ram pressure in the range of $113-194$ MPa when the meteor disintegrated. The required yield strength is $\gtrsim 20$ times higher than stony meteorites and $\gtrsim 2$ times larger than iron meteorites. The implied slowdown in the atmosphere suggests an initial speed of about $66.5 \; {\rm km~s^{-1}}$, strengthening the case for an interstellar origin of this meteor and making it an outlier relative to the velocity dispersion of local stars.
\end{abstract}



\section{Introduction}

Two interstellar objects have been detected so far in the solar system through their reflection of sunlight: `Oumuamua in 2017 \citep{2017Natur.552..378M}, and Borisov in 2019 \citep{2020NatAs...4...53G}. CNEOS\footnote{\url{https://cneos.jpl.nasa.gov/}} 2014-01-08, detected by U.S. Department of Defense (DoD) sensors through the light that it emitted as it burned up in the Earth's atmosphere off of the coast of Papua New Guinea in 2014, was determined to be an interstellar object in 2019 \citep{2019arXiv190407224S}, a conclusion that was confirmed by independent analysis conducted by the DoD in 2022 \citep{shaw_2022}. Recently, the light curve of CNEOS 2014-01-08 was released through the CNEOS database.\footnote{\url{https://cneos.jpl.nasa.gov/fireballs/lc/bolide.2014.008.170534.pdf}} Here, we investigate some basic implications of the light curve. 

\section{Light Curve Analysis}

First, we convert the optical power reported in the light curve to total power. Using Equation (1) from \cite{2002Natur.420..294B}, combined with the total optical energy for CNEOS 2014-01-08 of $3.1 \times 10^{17} \; \mathrm{erg}$ \citep{2019arXiv190407224S}, we find that the optical efficiency is $\sim 6.9\%$. The total power as a function of time is therefore simply the optical power as a function of time, divided by $6.9\%$.

\begin{figure}
  \centering
  \includegraphics[width=\linewidth]{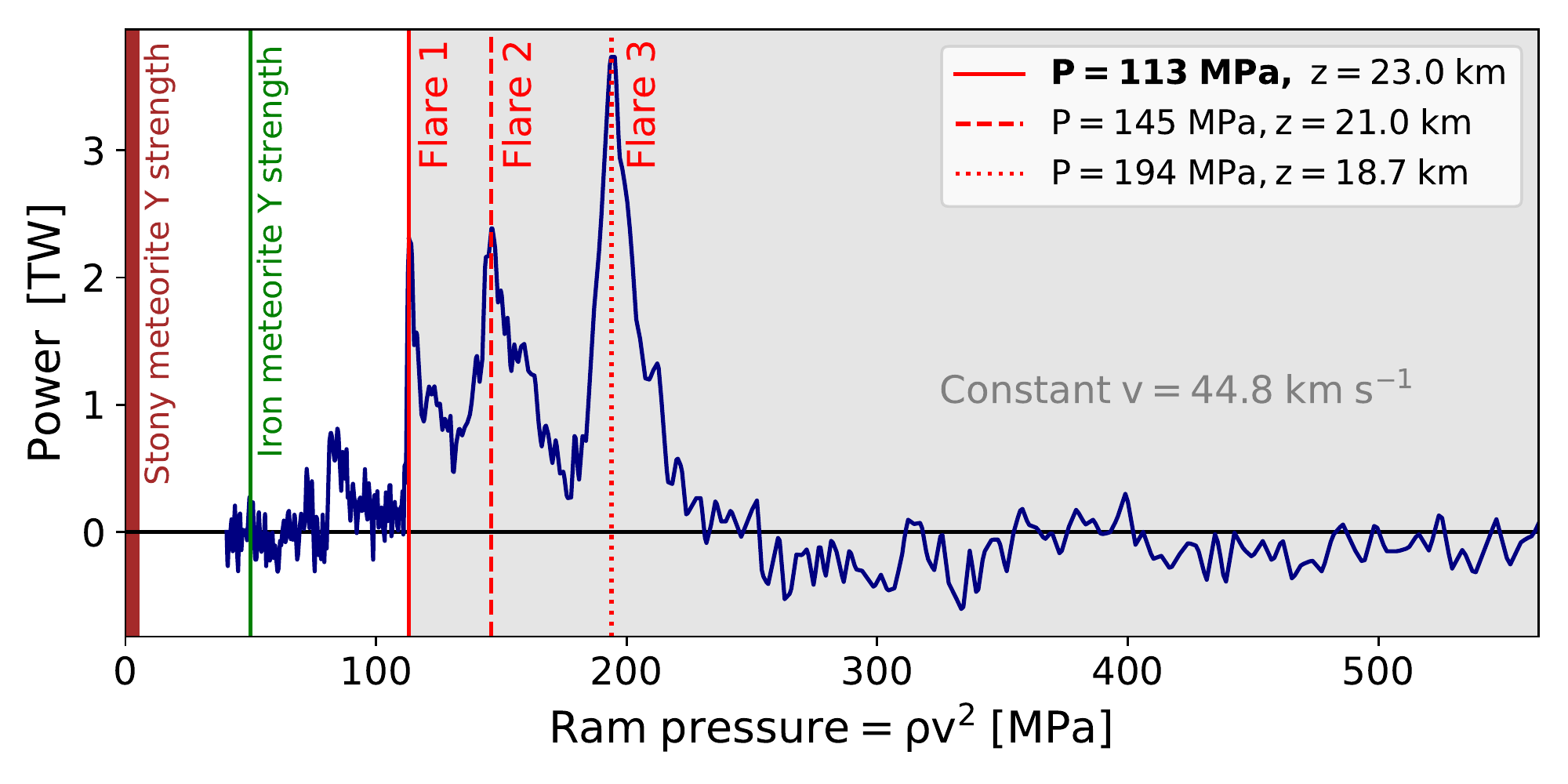}
    \caption{Total power released in the CNEOS 2014-01-08 fireball as a function of ram pressure, $\rho v^2$. Typical stony and iron meteorite yield strengths, $1 - 5 \mathrm{\; MPa}$ and $50 \mathrm{\; MPa}$ respectively, are indicated for convenience of comparison, and the three major flare events are labelled according to the order in which they occurred. We conservatively adopt a constant velocity between the flares. Note that 1 TW = $10^{19} \; \mathrm{erg \; s^{-1}}$ and 1 MPa = $10^7 \mathrm{\; dyne \; cm^{-2}}$.}
    \label{fig:p}
\end{figure}

Next, we derive the ram pressures, $\rho v^2$, corresponding to the three major explosion flares visible in the light curve, where $\rho$ is the ambient mass-density of air and $v$ is the meteor speed. We adopt a straight-line trajectory for CNEOS 2014-01-08 as it moves through the atmosphere at an angle of $\theta = 26.8^{\circ}$ relative to the ground \citep{2019RNAAS...3...68Z}. The velocity and altitude measurements reported by CNEOS, $v_{CNEOS} = 44.8 \mathrm{\; km \; s^{-1}}$ and $z_{CNEOS} = 18.7 \mathrm{\; km}$, correspond to peak brightness,\footnote{\url{https://cneos.jpl.nasa.gov/fireballs/intro.html}} or Flare 3 in Figure \ref{fig:p}. We conservatively adopt a constant speed of $v_{CNEOS} = 44.8 \mathrm{\; km \; s^{-1}}$ between the flares; deceleration due to object breakup between Flare 1 and Flare 3 would lead to greater ram pressures. The $\Delta t_{2,3} = 0.112 \; \mathrm{s}$ delay between Flares 2 and 3, and the $\Delta t_{1,3} = 0.213 \; \mathrm{s}$ delay between Flares 1 and 3, imply that Flares 2 and 3 occurred at $(v_{CNEOS} \Delta t_{2,3} \sin{\theta}) = 2.3 \mathrm{\; km}$ and $(v_{CNEOS} \Delta t_{1,3} \sin{\theta}) = 4.3 \mathrm{\; km}$ above the altitude at which Flare 3 occurred, respectively. This indicates that Flare 1 and Flare 2 occurred at altitudes of $z = 23.0 \mathrm{\; km}$ and $z = 21.0 \mathrm{\; km}$, respectively. 

We adopt the atmospheric density profile of $\rho(z) = \rho_0 \exp{(-z / H)}$, where $\rho_0 = 10^{-3} \mathrm{\; g \; cm^{-3}}$ is the sea-level atmospheric density and $H = 8 \mathrm{\; km}$ is the scale height of the Earth's atmosphere \citep{2005M&PS...40..817C}. The atmospheric densities at which Flare 1, Flare 2, and Flare 3 transpired are $\rho_1 = \rho(23.0 \mathrm{\; km}) = 5.64 \times 10^{-5} \; \mathrm{g \; cm^{-3}}$, $\rho_2 = \rho(21.0 \mathrm{\; km}) = 7.24 \times 10^{-5} \; \mathrm{g \; cm^{-3}}$, and $\rho_3 = \rho(18.7 \mathrm{\; km}) = 9.66 \times 10^{-5} \; \mathrm{g \; cm^{-3}}$. The resulting ram pressures for the three flares are $(\rho_1 v_{CNEOS}^2) = 113 \mathrm{\; MPa}$, $(\rho_2 v_{CNEOS}^2) = 145 \mathrm{\; MPa}$, and $(\rho_3 v_{CNEOS}^2) = 194 \mathrm{\; MPa}$, respectively. Figure \ref{fig:p} displays the flares in terms of total power as a function ram pressure.

\section{Implications for Yield Strength and Composition}

Breakup occurs when the yield strength of the impactor $Y_i$ is equivalent to the ram pressure: $Y_i = \rho v^2$ \citep{2005M&PS...40..817C}. CNEOS 2014-01-08's active phase is bracketed by a narrow range of ram pressures, $113-194$ MPa, which translates directly into the constraint on the original object's yield strength. Most conservatively, the yield strength of CNEOS 2014-01-08 was comparable to the ram pressure of Flare 1, $Y_i = (\rho_1 v_{CNEOS}^2) = 113 \mathrm{\; MPa}$.

Based on estimates for comets, carbonaceous, stony, and iron meteorites \citep{1993Natur.361...40C, 1993Natur.365..733S, 1995Icar..116..131S, 2001JMatS..36.1579P}, \cite{2005M&PS...40..817C} established an empirical strength-density relation for impactor density $\rho_i$ in the range $1 - 8 \mathrm{\; g \; cm^{-3}}$. The upper end of this range gives a yield strength of $Y_i \sim 50 \mathrm{\; MPa}$, corresponding to the strongest known class of meteorites, iron \citep{2001JMatS..36.1579P}. Iron meteorites are rare in the solar system, making up only $\sim 5\%$ of modern falls \citep{2006mess.book..869Z}. The CNEOS 2014-01-08 inferred yield strength of at least $Y_i = 113 \mathrm{\; MPa}$ exceeds the typical yield strength of iron meteorites by a factor of $\sim 2$. The observed yield strength is also inconsistent with stony meteorites, which exhibit a range of lower yield strengths by $1 - 2$ orders of magnitude \citep{2001JMatS..36.1579P, 2011M&PS...46.1525P}. Finally, the natural possibilities considered for `Oumuamua's composition are ruled out for CNEOS 2014-01-08 on the basis of insufficient strength, namely a nitrogen iceberg \citep{2021JGRE..12606706J, 2021JGRE..12606807D}, an $\mathrm{H_2}$ iceberg \citep{2020ApJ...896L...8S}, or a fluffy dust cloud \citep{2019ApJ...872L..32M, 2020ApJ...900L..22L}.

Additionally, CNEOS 2014-01-08 experienced slowdown between its atmospheric entry and detonation. We define a slowdown factor $f_s$, derived from Equation (8) in \cite{2005M&PS...40..817C},
\begin{equation}
    f_s(z) = \exp{\left(-\frac{3 \rho(z) C_D H}{4 \rho_i L_0 \sin \theta}\right)},
\end{equation}
where $C_D = 2$ is the drag coefficient and $L_0 = 2 \times (3 E / 2 \pi v_{CNEOS}^2 \rho_i)^{1/3}$ is the diameter of the object, where $E = (3.1 \times 10^{17}/ 6.9\%) \mathrm{\; erg}$ is the total explosion energy. The meteor's speed at an altitude $z$ above the range of breakup altitudes is then $v(z) \sim [f_s(z) v_{CNEOS} / f_s(z_{CNEOS})]$. Adopting a fiducial density of $\rho_i = 8 \mathrm{\; g \; cm^{-3}}$, corresponding to an iron composition and $L_0 \sim 0.5~{\rm m}$, this implies that the impactor's speed at the top of the atmosphere was at least $v(z \rightarrow \infty) = 66.5 \mathrm{\; km \; s^{-1}}$, which is $22 \mathrm{\; km \; s^{-1}}$ or $48 \%$ faster than $v_{CNEOS} = 44.8 \mathrm{\; km \; s^{-1}}$, the impact speed used to evaluate the orbit and determine the interstellar origin of CNEOS 2014-01-08 \citep{2019arXiv190407224S}. A lower value of $\rho_i$ would lead to greater slowdown. This increase in geocentric impact speed makes the interstellar origin of CNEOS 2014-01-08 clearer, and its motion relative to the local standard of rest even more anomalous \citep{2019arXiv190407224S}. 

%\pagebreak
\newpage
\section*{Acknowledgements}
%\vspace{0.1in} 
This work was supported in part by Harvard's {\it Black Hole Initiative}, which is funded by grants from JTF and GBMF. %\newline \newline

%\vspace*{0.8in} 



%% of citation expression problems, but it is  to clearly

\bibliography{bib}{}
\bibliographystyle{aasjournal}



\end{document}